\documentclass[aps,prl,twocolumn,groupedaddress,bibnotes]{revtex4}
\usepackage{graphicx,epsfig,subfigure,afterpage,bm}

\newcommand{\be}{\begin{equation}}
\newcommand{\ee}{\end{equation}}
\newcommand{\bea}{\begin{eqnarray}}
\newcommand{\eea}{\end{eqnarray}}
\newcommand{\gdot}{\dot{\gamma}}

\newcommand{\bw}{\begin{widetext}}
\newcommand{\ew}{\end{widetext}}

\newcommand{\vecv}[1]{\bm{{#1}}}
\newcommand{\tens}[1]{\bm{{#1}}}
\newcommand{\nablu}{{\bf \nabla}}

\begin{document}

\title{Viscoelastic Taylor-Couette instability of shear banded flow}
\author{Suzanne M. Fielding}
\email{suzanne.fielding@durham.ac.uk}
\affiliation{Department of Physics, University of Durham, Science Laboratories, South Road, Durham. DH1 3LE}
\date{\today}
\begin{abstract}
We study numerically shear banded flow in planar and curved Couette
geometries. Our aim is to explain two recent observations in shear
banding systems of roll cells stacked in the vorticity direction,
associated with an undulation of the interface between the
bands. Depending on the degree of cell curvature and on the material's
constitutive properties, we find either (i) an instability of the
interface between the bands driven by a jump in second normal stress
across it; or (ii) a bulk viscoelastic Taylor Couette instability in
the high shear band driven by a large first normal stress within it. Both
lead to roll cells and interfacial undulations, but with a different
signature in each case. Our work thereby suggests a different origin
for the roll cells in each of the recent experiments.
\end{abstract}
\pacs{ {a}, 
{b}.
     } 
\maketitle

The Taylor-Couette instability that arises when a Newtonian fluid is
sheared between concentric cylinders has a long history in classical
hydrodynamics, dating back to the groundbreaking paper of G.~I.~Taylor
in 1923~\cite{taylor1923}. The effect is inertial in origin: fluid is
forced centrifugally outward along the radial direction $r$, and
recirculates via roll cells stacked in the vorticity direction $z$. In
contrast, inertia is usually negligible in the non-Newtonian flows of
complex fluids. However the last two decades have seen considerable
interest in an inertialess, {\em viscoelastic} Taylor-Couette (VTC)
instability~\cite{larson-jfm-218-573-1990}, originating instead in the
hoop stresses (first normal stresses) that arise when a polymeric
fluid is sheared in a curved geometry. These squeeze fluid radially
inwards, again triggering an instability that leads to roll cells
stacked along $z$.

Another intensely studied flow phenomenon in complex fluids is that of
``shear banding''~\cite{reviews}, in which an initially homogeneous
shear flow gives way to a state of coexisting bands of unequal
viscosities and internal structuring, with layer normals in the
flow-gradient direction $r$. Close analogies exist between this
non-equilibrium transition and conventional equilibrium phase
coexistence, both kinetically and in the ultimate banded state. There
are also fundamental differences, {\it e.g.}, in the way the
coexistence state is selected in the absence of a free energy
minimisation principle~\cite{lu-prl-84-642-2000}.  Indeed, beyond the
basic observation of banding, an accumulating body of data shows that
many (perhaps most) shear banded flows show complicated
spatio-temporal patterns and dynamics~\cite{fielding2007a}. These are
often associated with roll cells stacked along $z$, in both curved
Couette~\cite{lerouge} and
planar~\cite{submitted} flow geometries, the origin of which remains
unclear.

In this Letter, we give the first theoretical evidence to suggest
that, in a curved flow, these rolls can arise via a mechanism in which
the high shear band, once formed, develops large enough first normal
stresses to trigger a further bulk patterning instability of the VTC
type within itself.  We further show that this bulk instability
disappears below a critical value of the cell curvature $q$, depending
in a quantifiable way on the fluid's constitutive properties, and
consistent with the known criterion~\cite{larson-jfm-218-573-1990}
that VTC instability occurs above a critical value of $\gdot
q^{1/2}$. At small curvatures, however, a different instability
emerges, driven by a jump in {\rm second} normal stress across the
interface between the bands~\cite{fielding2007b}. This also leads to
roll cells, but with identifiably different properties from those of
the bulk instability. By mapping in this way a phase diagram
containing these two separate instabilities -- a bulk VTC-like instability
of the high shear band, and an instability of the interface between
the bands -- we suggest different origins for the different
experimental observations of roll cells in
Refs.~\cite{lerouge,submitted}.

Our study thereby brings together three different hydrodynamic
instabilities in complex fluids: shear banding itself; instability of
an interface between bands; and, for the first time theoretically, a
bulk VTC-like instability of one band. We hope thereby to stimulate
further experiments, in a family of flow cells of different
curvatures, to verify (or otherwise) our findings with regards this
strikingly rich array of hydrodynamic instabilities.

{\em Governing equations ---} The generalised Navier--Stokes equation
for a viscoelastic material in a Newtonian solvent of viscosity $\eta$
and density $\rho$ is
\begin{equation} 
\label{eqn:NS} 
\rho(\partial_t +
  \vecv{v}.\nablu)\vecv{v} = \nablu.(\tens T -P\tens{I}) = \nablu .(\tens{\Sigma} + 2\eta\vecv{D}
  -P\tens{I}),
\end{equation} 
where $\vecv{v}(\vecv{r},t)$ is the velocity field and $\tens{D}$ is
the symmetric part of the velocity gradient tensor, $(\nablu
\vecv{v})_{\alpha\beta}\equiv \partial_\alpha v_\beta$. We assume zero
Reynolds' number $\rho=0$; and incompressibility $\partial_\alpha
v_\alpha=0$.  The quantity $\vecv{\Sigma}(\vecv{r},t)$ is the
viscoelastic contribution to the total stress $\vecv{T}(\vecv{r},t)$,
assumed to obey diffusive Johnson-Segalman (DJS)
dynamics~\cite{johnson}
%
%
\bea
\label{eqn:DJS}
(\partial_t
+\vecv{v}\cdot\nablu )\,\tens{\Sigma} 
&=& a(\tens{D}\cdot\tens{\Sigma}+\tens{\Sigma}\cdot\tens{D}) +
(\tens{\Sigma}\cdot\tens{\Omega} + \tens{\Omega}\cdot\tens{\Sigma})  \nonumber\\
& & + 2 G\tens{D}-\frac{\tens{\Sigma}}{\tau}+ \frac{\ell^2}{\tau }\nablu^2 
 \tens{\Sigma}.
\eea
Here $a$ is a slip parameter, which must lie in the range $-1\le a \le
1$; $G$ is a plateau modulus; $\tau$ is the viscoelastic relaxation
time; and $\tens{\Omega}$ is the antisymmetric part of the velocity
gradient tensor. The diffusive term $\nablu^2 \tens{\Sigma}$ is needed
to correctly capture the structure of the interface between the bands,
with a slightly diffusive interfacial thickness $O(l)$; and to ensure
unique selection of the shear stress at which banding
occurs~\cite{lu-prl-84-642-2000}.

\begin{figure}[tbp]
  \includegraphics[width=8.5cm]{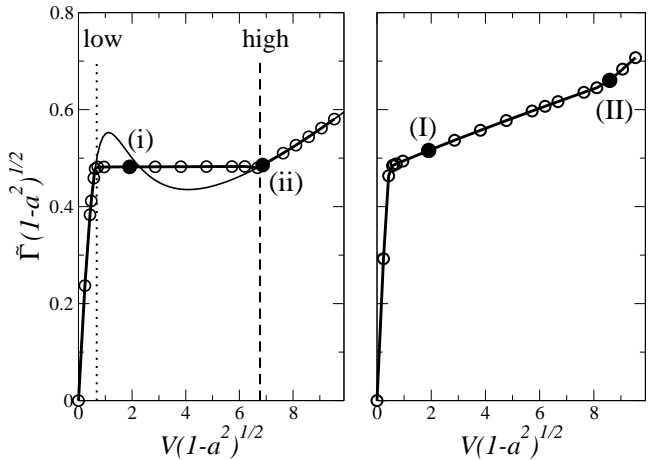}
  \caption{Bulk flow curve of 1D base state for $q=0.0$ {\bf (left)},
  $q=0.16$ {\bf (right)}.  Solid symbols (i), (ii), (I), (II) are for
  reference later in the manuscript. Styles of vertical lines
  indicating low and high shear bands are used consistently with
  Fig.~\ref{fig:phases}.}
\label{fig:flowCurves}
\end{figure}

{\em Geometry and boundary conditions ---} We study flow between
concentric cylinders of radii $r=\{R_1,R_2\}$, in cylindrical
coordinates $r,\theta,z$. The inner cylinder rotates at speed $V$; the
outer is fixed. We denote the cell curvature by $q=(R_1-R_2)/R_1$, the
flat limit of planar Couette flow corresponding to $R_1\to\infty$ at
fixed $R_2-R_1$, and so to $q\to 0$. The natural bulk rheological
variables are then the cylinder velocity $V$, which is the main
experimental control parameter; and the torque
$\tilde{\Gamma}=q^2\Gamma$, which we have rescaled by $q^2$ to remain
finite and equal to the shear stress $T_{xy}$ as $q\to 0$. We assume
invariance with respect to $\theta$ and study the model's dynamics in
the flow-gradient/vorticity plane $r-z$, which is the one most
commonly imaged experimentally. At the cylinders we assume boundary
conditions of zero flux normal to the wall
$\hat{\vec{n}}.\nablu\Sigma_{\alpha\beta}=0\;\forall\;\alpha,\beta$
for the viscoelastic stress, although others are
possible~\cite{adams2008}; and no slip or permeation for the
velocity. The height of the simulated region in the vorticity
direction is $L_z$, with periodic boundaries. We use units in which
$G=1,\tau=1$ and $R_2-R_1=1$. In these we choose parameter values
$\eta=0.05$, as suggested experimentally~\cite{lerouge,submitted}; a
cell height $L_z=2.0$; and an order of magnitude estimate
$l=O(10^{-3})$~\cite{fielding-epje-11-65-2003}.

\begin{figure}[tbp]
  \includegraphics[width=8.25cm]{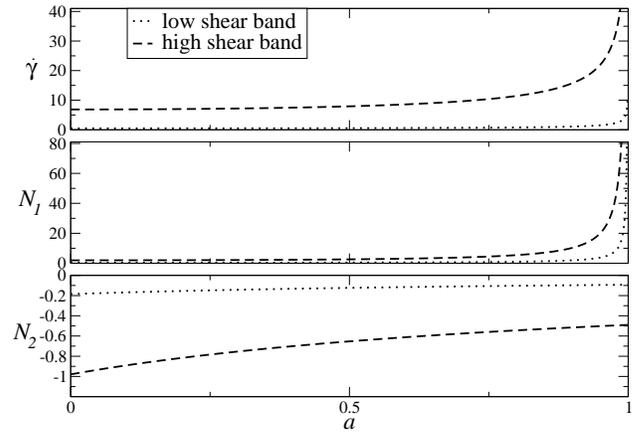}
  \caption{Shear rate and normal stresses in low and high shear
  bands {\it vs.} slip parameter $a$ (flat geometry, $q=0.0$).}
\label{fig:phases}
\end{figure}

{\em Numerics ---} We change variables from $(r,z)$ to $(p,z)$ where
$p=\ln(r/R_1)/\ln(R_2/R_1)$, so mapping our curved geometry onto an
effectively flat one, amenable to simulation on a regular rectangular
grid in which the only effect of curvature $q$ is to generate extra
source terms in the governing equations.
Our study holds for arbitrary values of $q$. Components of the
governing equations are then extracted, discretized on a grid
$(p_j,z_i)$, and evolved in time as described for the special case of
planar Couette flow, $q=0$, in Ref.~\cite{fielding2007b}. Results
below have timestep and grids $(Dt,Dp,Dz)=(0.0002,\alpha/512,1/512)$
with $\alpha=1$ (resp. $1/2$) for unbanded (resp. banded) base states,
convergence checked against smaller $(Dt,Dp,Dz)$.  Before generating
new data we checked that our code reproduces known results for (i)
the instability of an interface between shear bands in planar Couette
flow~\cite{fielding2007b};
(ii) 1D banded states in curved Couette flow~\cite{johnson};
(iii) dispersion relations for the onset of VTC instability in the
Oldroyd B model for small $q$~\cite{larson-jfm-218-573-1990};
(iv) neutral stability curves for this Oldroyd B VTC instability for
finite $q$~\cite{shaqfeh-jfm-235-285-1992}.

{\em Results: 1D banded states ---} First we discuss the results of
one-dimensional (1D) calculations that artificially assume
translational invariance in $z$, allowing structure only in the main
shear banding direction $r$. 

In the flat limit $q\to 0$, force balance dictates that the shear
stress $T_{xy}$ is uniform across the gap. Within the assumption of a
similarly homogeneous shear rate, the homogeneous constitutive relation
is given by $T_{xy}(\gdot)=\Sigma_{xy}(\gdot)+\eta\gdot$ where
$\Sigma_{xy}(\gdot)$ follows from solving Eqn.~\ref{eqn:DJS} subject
to invariance in time and space. See the thin solid line in
Fig.~\ref{fig:flowCurves} (left). For an applied shear rate in the
region of negative slope, homogeneous flow is linearly unstable and
the system separates into coexisting bands at a selected stress
$T_{\rm sel}\sqrt{1-a^2}=0.483$.  The steady state bulk flow curve
therefore shows a plateau in the banding regime (thick solid line in
Fig.~\ref{fig:flowCurves}, left), at this single value of the stress
for which a stationary interface can exist between bands. For non-zero
cell curvature the shear stress varies as $T_{xy}\sim 1/r^2$. This
forces the high shear band to reside next to the inner cylinder. As
this band expands outwards with increasing applied shear rate, the
overall torque increases to ensure that the interface between the
bands remains at the selected stress $T_{\rm sel}$. Consequently the
stress ``plateau'' in the banding regime acquires a non-zero,
curvature dependent slope: Fig.~\ref{fig:flowCurves}, right.

\begin{figure}[tbp]
  \includegraphics[width=8.5cm]{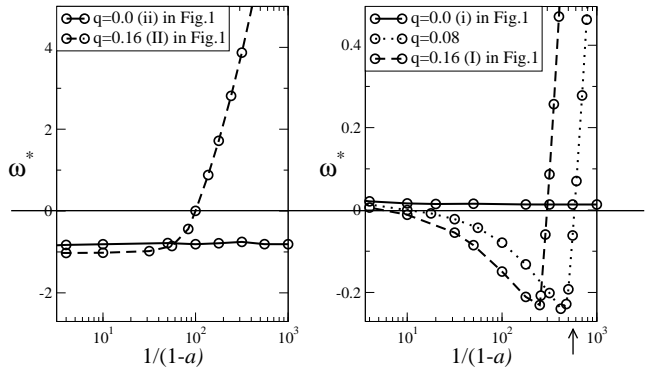}
  \caption{{\bf Left:} Maximum growth rate for an applied shear rate
    just into the high shear branch {\it vs.} the first normal stress
    scaling variable $1/(1-a)$, for two different cell curvatures:
    $(q,\gdot(1-a^2)^{1/2})=(0.16,8.58)$ and $(0.0,6.87)$.  {\bf
      Right:} Same for an applied shear rate $\gdot(1-a^2)^{1/2}=1.91$
    in the banding regime. States above the horizontal line are
    unstable, $\omega^*>0$. Value of $1/(1-a)$ in
    Fig.~\ref{fig:omega_with_q} is shown by an arrow.}
\label{fig:omega_with_a}
\end{figure}

In principle, these 1D base states must be calculated separately for
each set of values $(\eta,a)$. However we fix $\eta=0.05$ throughout,
and further exploit the fact that base states for all values of $a$
collapse onto a single master scaling state when expressed in terms of
the variables $\gdot\sqrt{1-a^2}$, $V\sqrt{1-a^2}$,
$\Gamma\sqrt{1-a^2}$, $\Sigma_{xy}\sqrt{1-a^2}$, $\Sigma_{xx}(1-a)$,
and $\Sigma_{yy}(1+a)$. Accordingly, the scaled flow curves in
Fig.~\ref{fig:flowCurves} represent all values of $a$ in the allowed
range $-1\le a\le 1$. Reading off the scaled shear rates of each band
(vertical lines in Fig.~\ref{fig:flowCurves}, left) we can then easily
extract their true shear rates as a function of $a$
(Fig.~\ref{fig:phases}, top). The corresponding first and second
normal stresses are also shown in Fig.~\ref{fig:phases} (for $q=0$,
but the trends are the same for a curved cell, $q\neq 0$). For $a\to
1$, the shear rate and first normal stress become large in the high
shear band. We therefore anticipate, and will show below, that for
sufficiently large cell curvatures $q$ this leads to a bulk
instability of the VTC kind in this band. Note that we are using $a$
as a parameter that de facto controls the strength of the normal
stresses, which is in reality set by a combination of material
properties such as wormlike micellar length and degree of
branching. We thereby hope to stimulate new experiments on shear
banding fluids with different constitutive properties.


\begin{figure}[tbp]
  \includegraphics[width=7.5cm]{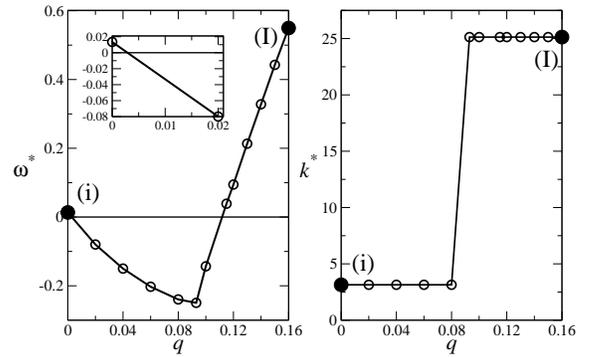}
  \caption{{\bf Left:} Maximum growth rate for a scaled applied shear
    rate $\gdot(1-a^2)^{1/2}=1.91$ in the banded regime, {\it vs.}
    cell curvature $q$. The slow interfacial instability at $q=0.0$
    (inset shows zoom near origin) is eliminated by curvature. At
    higher curvature, the high shear band displays a VTC-like
    instability. {\bf Right:} corresponding wavevector.  Symbols (i),
    (I) match those in Fig.~\ref{fig:flowCurves}, here for one fixed
    value of the first normal stress scaling variable $1/(1-a)=416$
    (arrow in Fig.~\ref{fig:omega_with_a}).}
\label{fig:omega_with_q}
\end{figure}

{\em Results: interfacial vs. bulk instability ---} We now move beyond
the assumption of 1D, performing 2D simulations in the $(r,z)$ plane.
In each run we take as our initial condition, or ``basic state'', a 1D
solution as just discussed, to which we add a 2D perturbation of tiny
amplitude. By monitoring the early time evolution of the Fourier modes
$\exp(ik z)\exp(\omega t)$ of this perturbation, we can extract the
dispersion relation $\omega(k)$ that characterises the linear
stability of the basic state. We then read off the maximum $\omega^*$
of this function. A value $\omega^*>0$ (resp. $\omega^*<0$) shows the
basic state to be linearly unstable (resp. stable) to 2D perturbations.

We start by considering a basic state comprising homogeneous flow on
the high shear branch in a flat geometry $q=0$, denoted (ii) in
Fig.~\ref{fig:flowCurves}. The corresponding dispersion relation
reveals this to be linearly stable, as expected: its peak $\omega^*<0$
(solid line in Fig.~\ref{fig:omega_with_a}, left). In contrast in a
curved device, denoted (II) in Fig.~\ref{fig:flowCurves}, the basic
state becomes linearly unstable when the first normal stress exceeds a
critical value, consistent with the onset of a bulk VTC-like
instability (dashed line in Fig.~\ref{fig:omega_with_a}, left). This
presence of a VTC-like instability in flow states that reside fully on
the high shear branch leads us to expect a bulk instability of the VTC
kind in the high shear band of a banded flow, in a curved cell $q >
0$. Accordingly, we now turn to the stability properties of a shear
banded basic state. We start with the flat case $q=0$ before turning
to the main situation of interest with $q>0$.

Shear banded flow in a flat geometry $q=0$ is already
known~\cite{fielding2007b} to show an instability driven by a jump in
second normal stress across the interface between the bands, leading
to undulations along the interface with wavelength $\lambda\approx 1$.
This is seen in Fig.~\ref{fig:omega_with_a} (solid line, right) for
the base state (i) in Fig.~\ref{fig:flowCurves}. The maximal growth
rate $\omega^*$ is weakly positive in the range $0<a<1$~\footnote{and
  becomes large as the second normal stress grows as $a\to -1$: not
  shown, and not expected in micelles}.

So far, then, we have brought together for the first time in the same
model three instabilities already documented separately in the
literature: shear banding itself~\cite{reviews}; a VTC-like instability of
a strongly sheared polymeric material with a large first normal stress
in a curved flow cell~\cite{larson-jfm-218-573-1990}; and the
instability in a flat geometry of an interface between shear bands
with respect to undulations with wavevector in the vorticity
direction~\cite{fielding2007b}, driven by the jump in second normal
stress across the interface.

Our most significant result, however, is to report the stability
properties of a shear banded state in two situations not previously
studied theoretically -- in a curved flow cell, and when the high
shear band has large first normal stress. We do so to demonstrate a
bulk VTC-like instability of the high shear band.

\begin{figure}[tbp]
\includegraphics[width=4cm,angle=90]{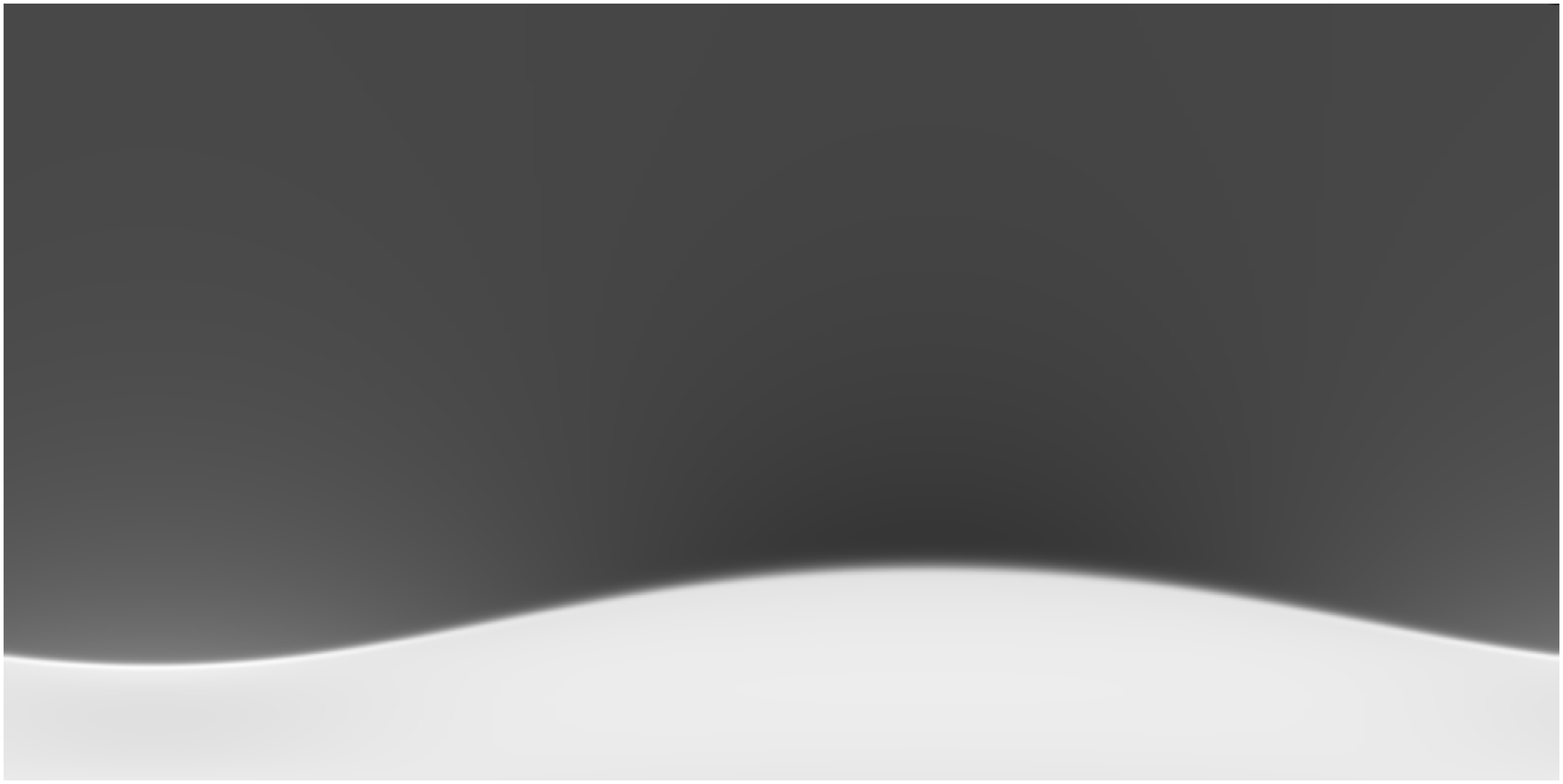}
\includegraphics[width=4cm,angle=90]{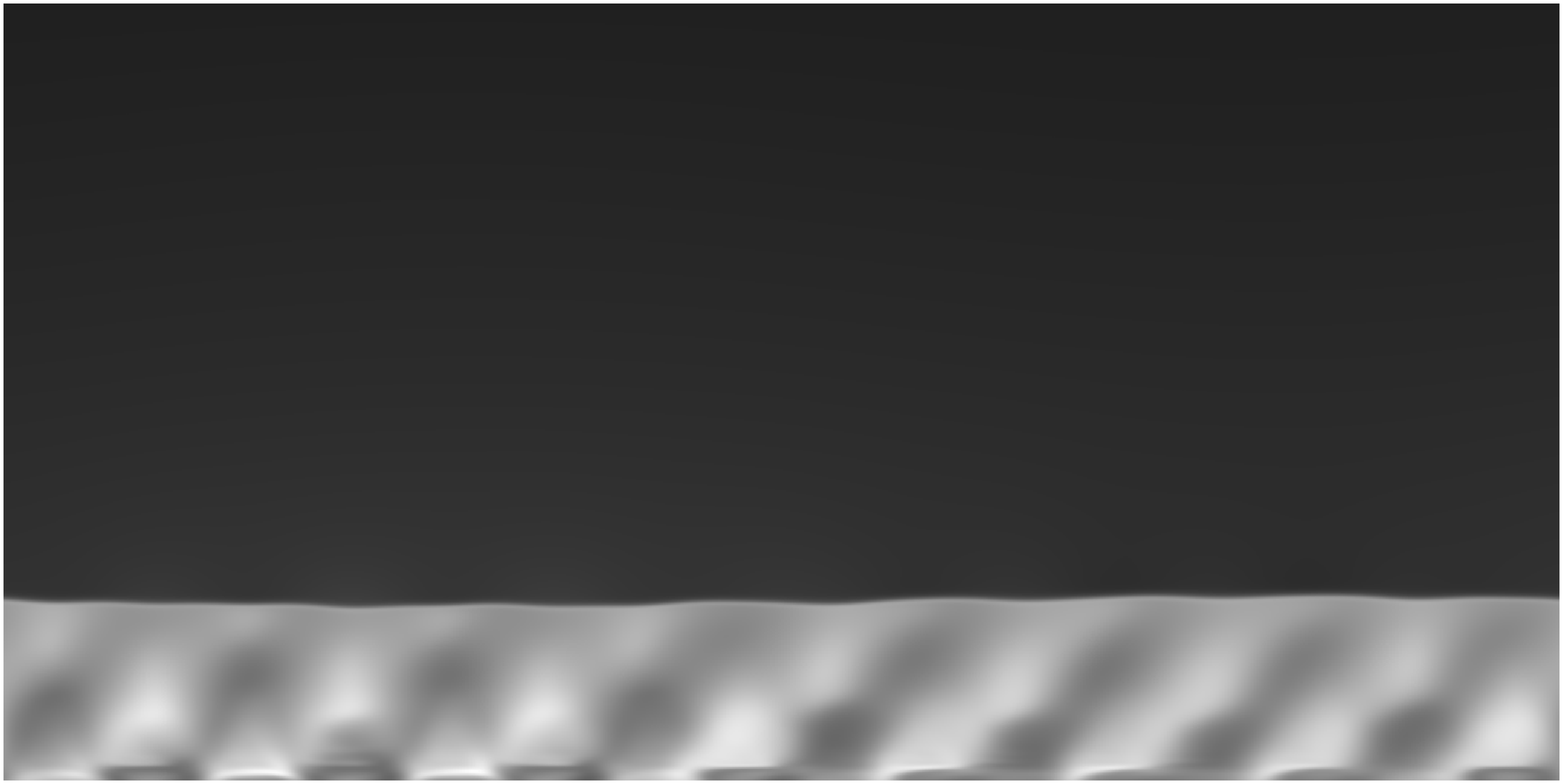}
\includegraphics[width=4cm,angle=90]{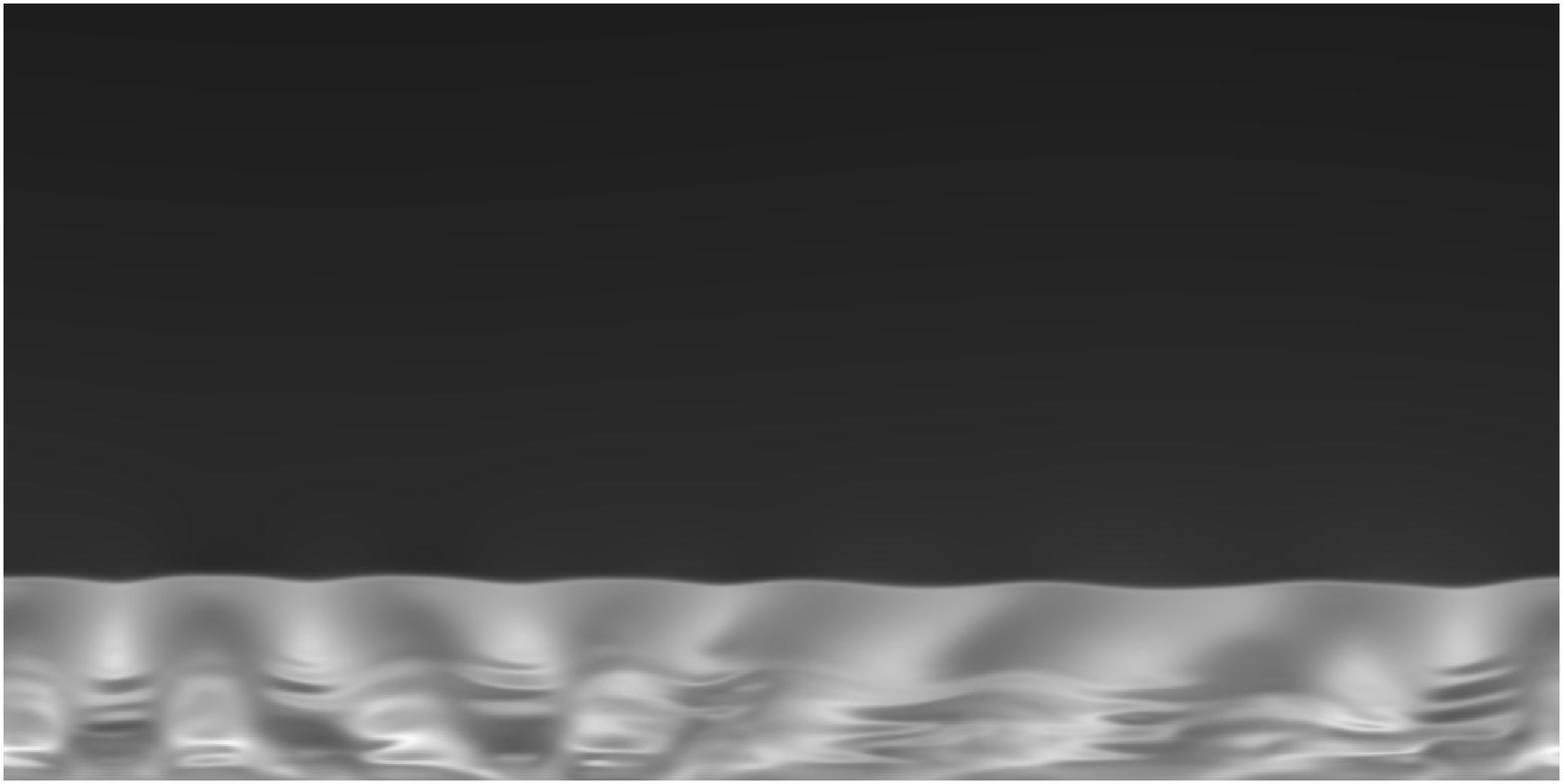}
\includegraphics[width=4cm,angle=90]{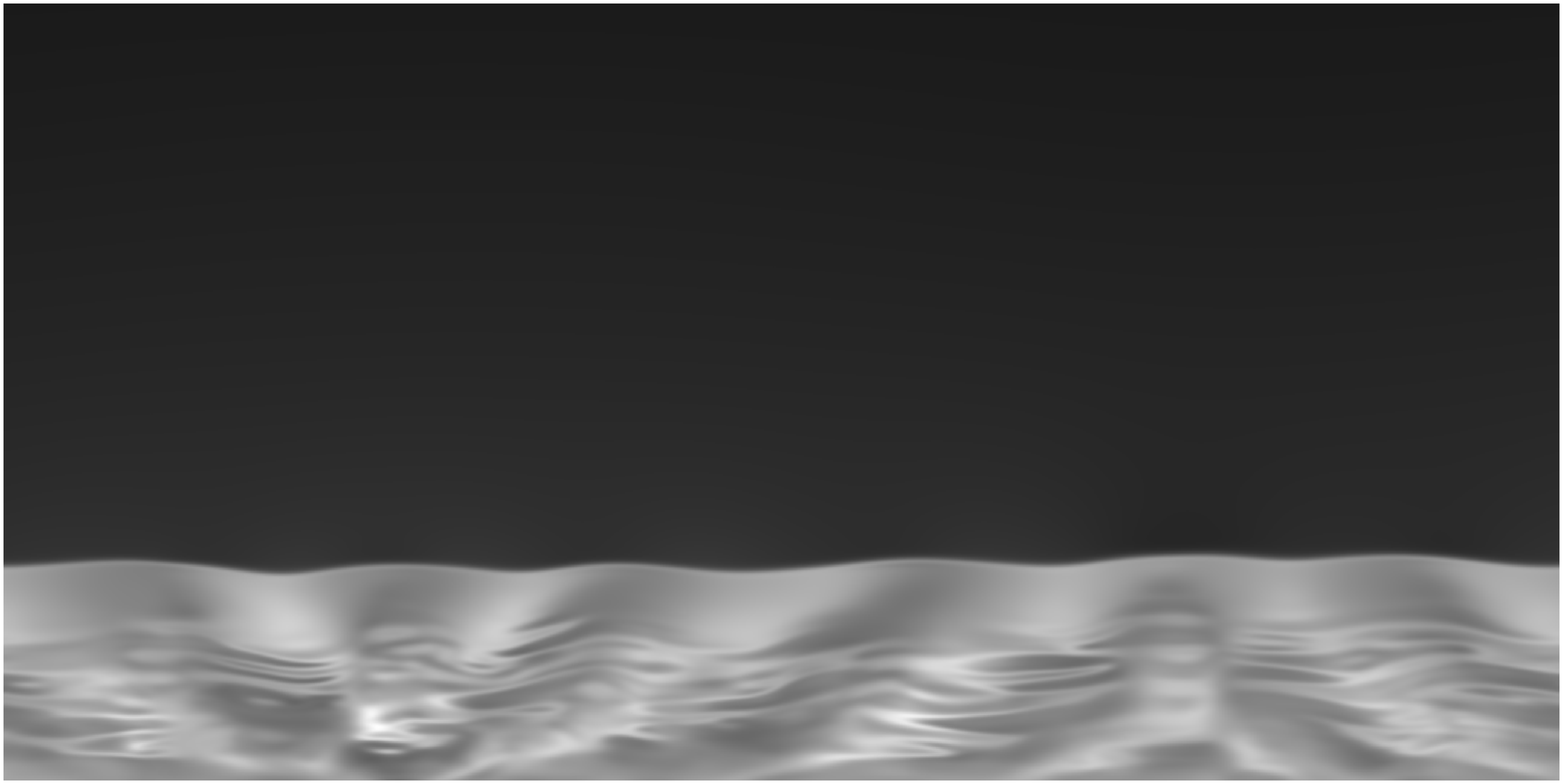}
\caption{Greyscale snapshots of $\Sigma_{xx}$ on the ultimate
  attractor in the $p-z$ (horizontal-vertical) plane for a shear rate
  in the banding regime $\gdot\sqrt{1-a^2}=1.91$. Left: interfacial
  instability in a flat cell $q=0$ with $1/(1-a)=1.43$. Others left to
  right: bulk instability of the VTC kind in the high shear band for a
  large value of the first normal stress scaling variable
  $1/(1-a)=416$, as in Fig.~\ref{fig:omega_with_q}, for
  curvatures $q=0.115,0.13,0.16$.}
\label{fig:state}
\end{figure}

As shown by the dotted and dashed lines in
Fig.~\ref{fig:omega_with_a}, right, curvature suppresses the
interfacial instability just discussed, dragging the weakly positive
mode for $q=0$ below the axis $\omega^*=0$ in the left hand part of
the plot. For large first normal stresses, though, we find a cross
over (interchange of dispersion maxima) to a different instability:
now of the bulk VTC kind in the high shear band. For a fixed value of
the first normal stress scaling variable $1/(1-a)$ the trend with cell
curvature is shown in Fig.~\ref{fig:omega_with_q}, left. Again the
interfacial instability (i) is suppressed by non-zero curvature $q>0$,
before the bulk VTC instability onsets in the high shear band at
$q\approx 0.112$. At the interchange, the wavelength of the fastest
growing mode switches (Fig.~\ref{fig:omega_with_q}, right) from
$\lambda \approx 1$, consistent with an interfacial instability, to
$\lambda \approx 1/8$, consistent~\cite{larson-jfm-218-573-1990} with
a VTC instability in a bulk flow phase of width $\approx 0.3$, for
this applied shear rate. Greyscale snapshots on the ultimate nonlinear
attractor for each kind of instability are shown in
Fig.~\ref{fig:state}. The stress signals of the rightmost of these
appear chaotic, consistent with the observation of complex roll cell
dynamics in Ref.~\cite{lerouge}. We defer to future work a detailed
study of the temporal roll cell dynamics in our simulations.

\begin{figure}[tbp]
  \includegraphics[width=7cm]{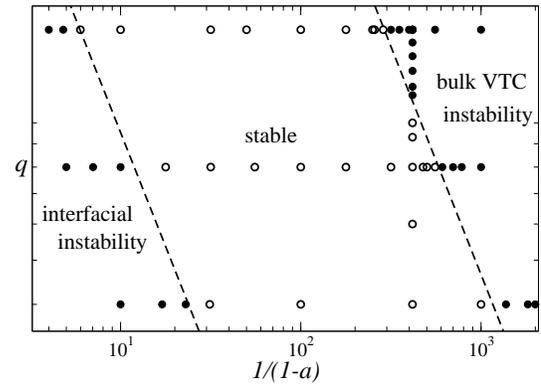}
  \caption{Phase diagram in the plane of curvature $q$ {\it vs.} first
  normal stress scaling variable $1/(1-a)$. Dashed lines show power
  $-1$ as a guide to the eye.}
\label{fig:phaseDiagram}
\end{figure}

{\em Summary and outlook ---} For large first normal stresses and cell
curvature we have demonstrated a bulk instability of the VTC kind in a
high shear band. The phase boundary (Fig.~\ref{fig:phaseDiagram})
scales as $q\sim (1-a)\sim N_1^{-1} \sim\gdot^{-2}$, consistent with
the criterion for VTC instability known for bulk unbanded
flow~\cite{larson-jfm-218-573-1990}. In contrast, for small curvatures
and first normal stresses we find an undulatory instability of the
interface between the bands as reported in Ref.~\cite{fielding2007b}
for $q=0$. An important additional contribution of this work has been
to show this interfacial instability to be suppressed by cell
curvature, our data suggesting the same threshold scaling $q\sim
(1-a)$ as for VTC.

The aim of this Letter has been to shed light on two different recent
experimental observations of roll cells in shear banded
flows~\cite{lerouge,submitted}. Nghe et al.~\cite{submitted}
demonstrated an instability leading to rolls stacked along the
vorticity direction $z$ in a rectilinear microchannel. The lack of
cell curvature suggests these experiments to correspond to (the
pressure driven equivalent of~\cite{accepted}) our (boundary driven)
calculations for $q=0$ in Fig.~\ref{fig:phaseDiagram}, and so to a
linear interfacial instability. Indeed, it has long been known that
flows with parallel streamlines are linearly stable with respect to
bulk perturbations, although a nonlinear (subcritical) instability
cannot be ruled out~\cite{morozov-prl-95--2005}.

Lerouge and coworkers~\cite{lerouge}
demonstrated roll cells stacked along $z$ in a Couette cell of
curvature $q=O(0.1)$. They further estimated their high shear band to
satisfy the criterion for bulk VTC instability, which would correspond
to the top-right of our Fig.~\ref{fig:phaseDiagram}. Indeed, the
complex dynamics of
Ref.~\cite{lerouge} do perhaps
suggest VTC, although the wavelength $\approx 1$ seems more consistent
with interfacial instability. To resolve this issue fully, it would be
extremely interesting to perform a series of experiments scanning
right across Fig.~\ref{fig:phaseDiagram}. Most immediately, this could
be achieved by using a family of flow cells of different $q$. While
the stability gap in Fig.~\ref{fig:phaseDiagram} apparently covers an
impractical range of curvatures, we would expect smaller values of the
high shear viscosity $\eta$ (not uncommon experimentally but difficult
to access numerically) to narrow this gap significantly, such that it
indeed has the potential to be spanned by a family of cells.

Other open questions include the interaction these roll cells with
unstable modes of wavevector in the flow
direction~\cite{fielding-prl-95--2005}; and the effect of stick-slip
dynamics at the wall on this rich array of hydrodynamic phenomena.

{\it Acknowledgements} The author thanks Mike Cates, Ron Larson and Peter
Olmsted for discussions; and EPSRC (EP/E5336X/1) for funding.


\end{document}